\documentclass{article}

\usepackage{arxiv}

\usepackage[utf8]{inputenc} 
\usepackage[T1]{fontenc}    
\usepackage{hyperref}       
\usepackage{url}            
\usepackage{booktabs}       
\usepackage{amsfonts}       
\usepackage{nicefrac}       
\usepackage{microtype}      
\usepackage{lipsum}		
\usepackage{graphicx}
\usepackage{natbib}
\usepackage{doi}

\title{Hydra: Brokering Cloud and HPC Resources to Support the Execution of Heterogeneous Workloads at Scale}

\date{}

\author{\href{https://orcid.org/0000-0001-7491-4946}
        {\includegraphics[scale=0.06]{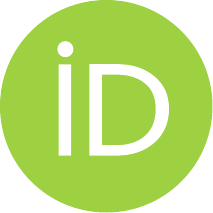}
        \hspace{1mm}
        Aymen Alsaadi}\\
	Department of Electrical and Computer Engineering\\
	Rutgers, the State University of New Jersey\\
	Piscataway, NJ 08854, USA \\
	\texttt{aymen.alsaadi@rutgers.edu} \\
	\And
	\href{https://orcid.org/0000-0000-0000-0000}
        {\includegraphics[scale=0.06]{orcid.pdf}
        \hspace{1mm}Shantenu Jha} \\
	Department of Electrical and Computer Engineering\\
	Rutgers, the State University of New Jersey\\
	Piscataway, NJ 08854, USA \\
	\texttt{shantenu.jha@rutgers.edu} \\
        Brookhaven National Laboratory\\
        Upton, NY, 11973, USA\\
        \texttt{shantenu@bnl.gov}
	\And
	\href{https://orcid.org/0000-0003-0527-1435}
        {\includegraphics[scale=0.06]{orcid.pdf}
        \hspace{1mm}Matteo Turilli} \\
	Department of Electrical and Computer Engineering\\
	Rutgers, the State University of New Jersey\\
	Piscataway, NJ 08854, USA \\
	\texttt{matteo.turilli@rutgers.edu}\\
        Brookhaven National Laboratory\\
        Upton, NY, 11973, USA\\
        \texttt{mturilli@bnl.gov}
}

\renewcommand{\headeright}{}
\renewcommand{\undertitle}{}
\renewcommand{\shorttitle}{Hydra: Brokering Cloud and HPC Resources to Support the Execution of Heterogeneous Workloads at Scale}

\hypersetup{
pdftitle={A template for the arxiv style},
pdfsubject={q-bio.NC, q-bio.QM},
pdfauthor={Aymen Alsaadi, Shantenu Jha, Matteo Turilli},
pdfkeywords={cloud computing · high performance computing · brokering},
}

\begin{document}
\maketitle

\begin{abstract}
Scientific discovery increasingly depends on middleware that enables the execution of heterogeneous workflows on heterogeneous platforms. One of the main challenges is to design software components that integrate within the existing ecosystem to enable scale and performance across cloud and high-performance computing (HPC) platforms. Researchers are met with a varied computing landscape, which includes services available on commercial cloud platforms, data and network capabilities specifically designed for scientific discovery on government-sponsored cloud platforms, and scale and performance on HPC platforms. We present Hydra, an intra/cross cloud/HPC brokering system capable of concurrently acquiring resources from commercial/private cloud and HPC platforms and managing the execution of heterogeneous workflow applications on those resources. This paper offers four main contributions: (1) the design of brokering capabilities in the presence of task, platform, resource, and middleware heterogeneity; (2) a reference implementation of that design with Hydra; (3) an experimental characterization of Hydra's overheads and strong/weak scaling with heterogeneous workloads and platforms; and (4) the implementation of a workflow that models sea-rise with Hydra and its scaling on cloud and HPC platforms.

\end{abstract}


\section{Introduction}\label{sec:intro}
Scientific discovery increasingly relies on workflow applications to perform experiments at unprecedented scale~\cite{ben2020workflows} and across a growing number and types of computing platforms. Those experiments execute workflows with heterogeneous tasks on heterogeneous resources~\cite{badia2017workflows}. Tasks may process large amounts of data and execute simulations, data analyses, and machine learning methods using cores, GPU, and AI accelerators on cloud and high-performance computing (HPC) platforms at different scales~\cite{da2021community}. Heterogeneity and scale pose unprecedented challenges for the middleware that supports the execution of workflow applications. While scientists can count on hundreds of workflow systems
, existing end-to-end solutions tend to be bespoke, specific to a programming model, type of workflow application, and platform and resource. Further, existing solutions are not specifically designed for performance at scale and often do not offer the required robustness, resilience, and efficiency~\cite{da2017characterization}.

Middleware designed to support modern scientific workflows should be building blocks, i.e., integrate with existing software systems without requiring a new code base~\cite{turilli2019middleware}. Middleware should not be yet another bespoke end-to-end solution; instead, it should contribute to a composable ecosystem that users can leverage to develop workflow solutions. Within that ecosystem, general purpose and extensible brokering capabilities have become critical. Workflow applications need to acquire and manage heterogeneous resources across various platforms and then manage the execution of heterogeneous tasks over those resources. That allows users to take advantage of the capabilities offered by diverse platforms. Specifically, scientific workflows require the data and network capabilities of publicly funded cloud and HPC platforms, the variety of services offered by commercial cloud platforms, and the scale supported by leadership-class HPC platforms.

This paper introduces Hydra, a brokering system that enables resource management and heterogeneous task execution across commercial and private cloud and HPC platforms. Hydra is a middleware component implemented in Python and designed to interface with existing workflow and runtime systems. Hydra is agnostic towards the application programming model used to implement the workflow application and uses dedicated connectors to concurrently interface with commercial and private service interfaces. When available, connectors support different types of services within each platform, enabling users to acquire resources at different levels of abstraction, e.g., via a batch system or a container. As a broker, Hydra does not offer an end-to-end solution for implementing workflow applications. For example, Hydra does not provide orchestration or workflow management capabilities. Instead, Hydra offers brokering capabilities to integrate and augment the existing scientific middleware ecosystem.

This paper offers four main contributions: (1) the design of brokering capabilities with task, platform, resource, and middleware heterogeneity; (2) Hydra, a reference implementation of that design; (3) an experimental characterization of Hydra's overheads and strong/weak scaling with heterogeneous workloads and platforms; and (4) Hydra's implementation and scaling of a workflow that models sea-rise on cloud and HPC platforms. Hydra enables executing workflows with diverse requirements: single/multi core/GPU/node and MPI/OpenMP\@. User-specified brokering policies determine whether those tasks are implemented as executables or containers and executed on cloud or HPC resources.

\section{Related Work}\label{sec:related}
\label{sec:others}

Several cloud brokers offer access to commercial cloud providers like Amazon Web Services
(AWS), Microsoft Azure
, and Google Cloud
. However, few brokers also offer access to publicly funded cloud providers, such as the platforms sponsored by the National Science Foundation (NSF), e.g., Jetstream2
and Chameleon
, and even fewer offer also concurrent access to HPC platforms.

CloudBridge~\cite{Goonasekera2016CloudBridge} is a Python Library that supports cross-cloud access via a unified and extensible application programming interface (API). CloudBridge supports access to commercial and private cloud providers such as AWS and NSF Chameleon. While CloudBridge unifies multiple Cloud APIs under a single user interface, it lacks resource and workload management capabilities.

CloudMesh~\cite{las2017cloudmesh} is a command line toolkit that enables users to access hybrid multi-cloud environments. CloudMesh enables access to AWS, Azure, Google Cloud, and NSF Chameleon. CloudMesh wraps these providers' APIs, enabling users to access them via a unified interface. However, CloudMesh does not offer additional capabilities, such as brokering and execution management.

EasyCloud~\cite{anglano2020easycloud} supports AWS, Azure, Google Cloud, and NSF Chameleon. EasyCloud manages and monitors multiple Virtual Machines (VMs), enabling real-time decision-making to prevent failure scenarios. However, EasyCloud does not provide workload-level monitoring and management capabilities, requiring additional capabilities when executing workflows across different cloud providers.

CompatibleOne~\cite{comaptible_2014_yangui} enables using different services managed by an OpenStack infrastructure. CompatibleOne's design is based on the Open Cloud Computing Interface (OCCI), offering ease of portability and extensibility. CompatibleOne implements a client/server architecture where the end-user machines are the clients, and cloud providers act as servers to deploy their agents and services. That architecture requires additional resources for component deployment and imposes communication and coordination overheads between the clients and the server.

Cloud Brokering~\cite{qos_cloudbroker_Quarati_2015} supports cloud providers that use OpenStack, Eucalyptus, and OpenNebula, and several commercial providers. Like CompatibleOne, Cloud Brokering uses OCCI and a client/server architecture. In addition to the already mentioned limitations, the Cloud Brokering's DCI-Bridge introduces a single point of failure and, possibly, a performance bottleneck.

BeeFlow~\cite{chen2018beeflow} is a cloud-HPC workflow manager that supports orchestrating hybrid workflows on HPC and cloud environments. BeeFlow supports commercial cloud providers with Container As A Service (CaaS) interfaces. In addition, BeeFlow supports HPC platforms with a build and execution environment (BEE) based on a containerization environment. However, BeeFlow does not support NSF-based platforms and offers only container-based workflow execution.

Domain-specific cloud brokers offer solutions to specific use cases. For example, EVOp~\cite{evop_Freer_2011} is a cloud broker designed for environmental use cases, HealthyBroker~\cite{kurdi2019healthybroker} serves patient-client use cases, and NLUBroker\cite{nlu_broker_2019} supports AI use cases that require natural language processing. These brokers support specific application types, which makes extending their design to general use cases difficult.

More general-purpose cloud brokers adopt better information flow and coordination to serve diverse applications. Ref.~\cite{WANG2015129} proposes a three-tier cloud broker architecture, which allows for service recommendation and combination. Schlouder~\cite{schlouder_2017} proposes a cloud broker, mainly compatible with open source tools---e.g., OpenStack---which offers seamless communications across different cloud providers. Nonetheless, both cloud brokers suffer from the limitations of a tightly coupled design, such as difficult integration with third-party middleware and limited scalability.

Other Cloud brokers~\cite{abdelbaky2015docker, Kalyani2018BuildingBO,
jrad_multi_cloud_workflows_2013, YANG20121158, cloudsim_2011_calheiros} present designs and architectures that offer new features and functionalities but are either no longer available or maintained.

\section{Hydra Cloud Brokering System}\label{sec:Hydra}
Hydra implements general-purpose brokering of heterogeneous services across commercial and private cloud and HPC platforms, addressing the main shortcomings of the existing solutions introduced in \S\ref{sec:related}. Specifically, Hydra can concurrently provision and monitor multiple commercial and NSF-sponsored cloud/HPC resources, and then broker, monitor, and trace heterogeneous workloads on those resources. Further, Hydra offers a simple and concise Python Application Programming Interface (API) that can accommodate multiple programming models, depending on whether the user codes directly against Hydra's API or integrates it with third-party software systems. Finally, Hydra's design uses stand-alone components that promote reusability, facilitate maintenance, and allow new services to be added without rewriting its code base.

Hydra implements the requirements of resource brokering: service management, integration, security, monitoring, and reporting~\cite{venkateswaran2020new}. Hydra standardizes the access and management of multiple resource providers (commercial and open source) by aggregating their APIs into a unified interface. That allows users to interact with multiple platforms while hiding their specific API and configuration systems. Further, Hydra offers intermediation capabilities by acting on behalf of the user to request, instantiate, and monitor resources while managing and monitoring the execution of user applications on different providers. These capabilities allow for the scalable execution of various scientific workloads.

\subsection{Architecture}\label{ssec:design}

\begin{figure}
    \centering
    \includegraphics[width=0.75\textwidth]{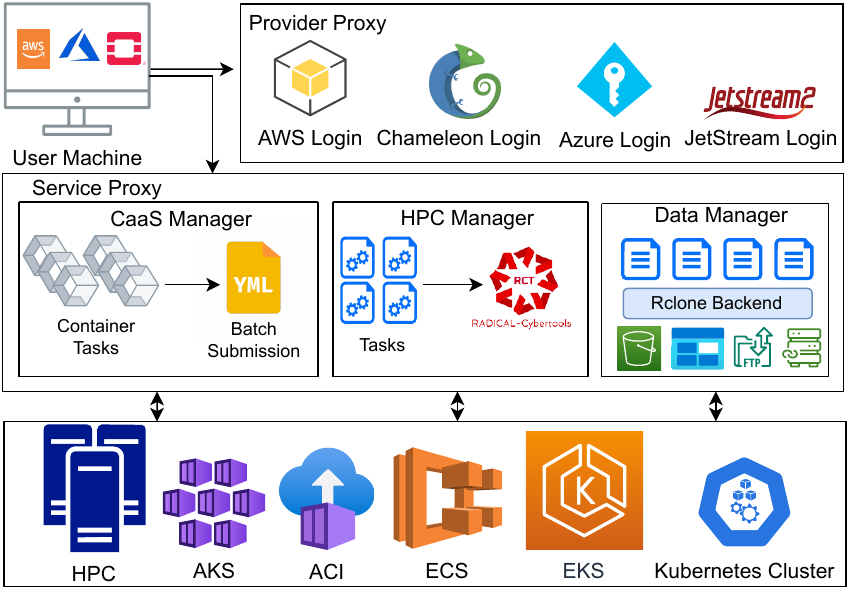}
	
    \caption{Hydra Architecture.}\label{fig:Hydra_architecture}
\end{figure}

Hydra's architecture has two main components (Fig.~\ref{fig:Hydra_architecture}): Provider Proxy and Service Proxy. Provider Proxy collects information about the user and the provider interfaces, verifying the user's credentials to guarantee the successful startup of Hydra's engine and services. Service Proxy implements Hydra's brokering capabilities, exposing service managers to concurrently interact with multiple cloud services and HPC batch systems. Further, the Service Proxy maps workloads to each service manager and monitors each manager and workload at runtime.

Currently, the Service Proxy has three subcomponents---CaaS Manager, HPC Manager, and Data Manager---but it exposes a private interface to add new managers like, for example, a function as a Service manager. The CaaS Manager manages container services and partitioning, submitting, monitoring, and tracing workloads. The CaaS Manager supports the container services of all the major commercial cloud platforms and can deploy and manage multi-node Kubernetes clusters on commercial and private cloud providers.

Hydra's HPC Manager supports multiple connectors, each designed to utilize the interface of an HPC middleware component. Currently, Hydra implements a connector for RADICAL-Pilot~\cite{Merzky2021DesignAP}, a pilot-enabled runtime system that allows the acquisition of HPC resources and the execution of heterogeneous workloads on them.

The Data Manager enables inter- and cross-cloud/HPC data management capabilities to support user workloads and data requirements. The manager implements data operations like copy, move, link, delete, and list, both locally and remotely. As with the HPC Manager, the Data Manager supports integration with different data management services as backends and exposes their operations via a unified API. Users can embed advanced data strategies in their applications, e.g., triggering data staging across sites or within a site with multiple storage systems. In the future, Hydra will expose methods to cache and prefetch data, hiding the complexity of the communication and coordination protocols from the user.

\subsection{Implementation}

Hydra is implemented as a Python module, and its API has four classes: Provider, Service, Resource, and Task. Provider exposes the \texttt{proxy} method to instantiate a Provide Proxy component, load the credentials and cloud provider configuration, and perform the credential validations. Service Proxy exposes methods to interface with cloud services and HPC batch systems. Resource exposes methods for each supported cloud and HPC provider, allowing users to specify the type of service they want to use, the amount of resources for each service, and all the other resource properties required by each provider and service. Task extends the class~\texttt{conccurent.Future}, allowing users to specify their workload properties and assign them to specific providers. Hydra's tasks map to regular executables, cloud pods, or containers, enabling users to set, for example, the task provider, container path, memory per task, and CPU/GPU units per task. Each task object also holds information about its current/final state and tracing events.

Hydra's CaaS Manager can instantiate new clusters on each cloud provider from the requirements specified via the \texttt{resource.VM} object. Based on the available resources of each cluster, the CaaS Manager partitions the given workload into batches that fit the available resources. Once the requested resources and services---including the data services available on the target cloud platform---are ready, the CaaS Manager submits the tasks to the service interface of each provider in a single batch. That reduces the communication between Hydra and the provider, reducing Hydra's overheads and increasing its throughput.

The CaaS Manager traces the concurrent execution of all tasks until they are in a final state, i.e., \texttt{done}, \texttt{canceled}, or \texttt{failed}. Due to the cost associated with moving data from the cloud service to the user machine, task traces and outputs are not stored in the task object unless specified by the user. The HPC Manager uses the RADICAL-Pilot connector to bulk-submit resource requirements and task descriptions. Like the CaaS Manager, the HPC Manager monitors the submitted tasks via RADICAL-Pilot and, if requested by the users, retrieves the traces of the task executions. Hydra managers ensure graceful terminations of all the instantiated resources upon completion of the whole workload or, when configured by the user, upon failure of one or more tasks.

\section{Exemplar Use Case}\label{sec:usecases}

We introduce FACTS (Framework for Assessing Changes To Sea-level) as an exemplar use case, showing how Hydra's capabilities allow concurrently running a real-life workflow on multi-node Kubernetes clusters and HPC resources. In \S\ref{sec:exp}, we present Hydra's performance characterization when implementing and executing one of FACTS' workflow.

FACTS is a Python tool for projecting sea-level rise~\cite{kopp2023facts_code}, which offers a modular platform for characterizing parametric and structural uncertainty in future global, relative, and extreme sea-level changes. FACTS consists of modules to simulate the different processes that contribute to sea-level changes~\cite{kopp2023framework}. Each module is an independent workflow that can currently be executed on a container or scaled up only on HPC platforms via RADICAL-EnsembleToolkit (EnTK)~\cite{balasubramanian2018harnessing}.

Modules under development use various AI/ML methods~\cite{van2023variational} that will require cloud environments to scale. Currently, FACTS requires $\sim$21 GB of data, but that will grow 10/100-fold, requiring cloud platforms that allow for long-term storage of various datasets. Finally, with the planned contribution of third-party modules to FACTS from diverse domain scientists, executing FACTS on a cloud platform could offer a stable, predictable, and testable development and production environment.

Supporting the execution of FACTS on commercial/private cloud and HPC platforms poses three main requirements: (1) FACTS modules must be containerized, separating data and compute capabilities while offering execution environments with the capabilities required by ML/AI, simulation and analysis tasks; (2) scaling the concurrent execution of modules requires an elastic runtime environment, alongside suitable workflow management capabilities; and (4) data capabilities must be available to access datasets, possibly via dedicated service interfaces. Hydra addresses all FACTS requirements to run on the cloud efficiently while continuing to run the current FACTS implementation on HPC platforms.

Note that FACTS is an exemplary use case and that we did not explicitly design Hydra to support it. As seen in \S\ref{ssec:design}, Hydra is a general-purpose cloud and HPC broker specifically designed to support diverse programming models and interfaces. Hydra is agnostic towards the type of science performed by the workload/workflows it brokers.

\section{Performance Characterization}\label{sec:exp}

Table~\ref{table:exp_setup} shows the setup of our experiments on diverse cloud and HPC platforms. We use the NSF Chameleon~\cite{keahey2020lessons} and Jetstream2~\cite{jetstream2_2021} cloud providers, and the ACCESS Bridges2~\cite{brown2021bridges} HPC platform. Chameleon is an experimental cloud platform with 550 nodes and 5 PB of storage, Jetstream2 is a production-grade cloud platform with 448 compute nodes and 17.2 PB of storage, and Bridges2 is an HPC + AI + Data cluster with 603 compute nodes and 10 PB of storage. We use Amazon Web Services (AWS) and Microsoft Azure as commercial cloud providers.

\begin{table*}[h]
       \hspace{0.5mm}
	\label{table:exp_setup}
	\centering
	\begin{tabular}{llllllll}
	\toprule
	ID              &
	Exp. Type       &
	Workload Type   &
	Plat. Type      &
	No. Tasks       &
	Task Type       &
	Nodes Per Run   &
	Total CPUs      \\
	\midrule
	1               &
	P-PR            &
	HOM             &
	Cloud           &
	[4,8,16]K       &
	CON       &
	1               &
	[4--16]         \\
	2               &
	C-PR            &
	HOM             &
	Cloud           &
	[16,32,64]K     &
	CON &
	1               &
	16              \\
	3-A             &
	C-PL            &
	HOM             &
	Cloud-HPC       &
	[20,40,80]K     &
	CON       &
	1               &
	16              \\
	3-B             &
	C-PL            &
	HET             &
	Cloud-HPC       &
	10,240          &
	CON, EXEC &
	[2,4,6]         &
	[4--128]        \\
	4               &
	FACTS           &
	HET             &
	Cloud-HPC       &
	200-3200        &
	CON, EXEC &
	[1,2,4,8,16]    &
	[16--256]       \\
	\bottomrule
	\end{tabular}
\caption{Setup of Experiment 1, 2, 3, and 4. P-PR = Per Provider; C-PR = Cross Providers; C-PL = Cross Platform; HOM = homogeneous; HET = heterogeneous; CON = container; EXEC = executable. Task executables: 1--3A = \texttt{noop}; 3B = \texttt{sleep}; 4 = \emph{pre-processing}, \emph{fitting}, \emph{projecting} and \emph{post-processing}.}

\end{table*}
\begin{figure*}[h]
    \centering
    \includegraphics[width=1\textwidth]{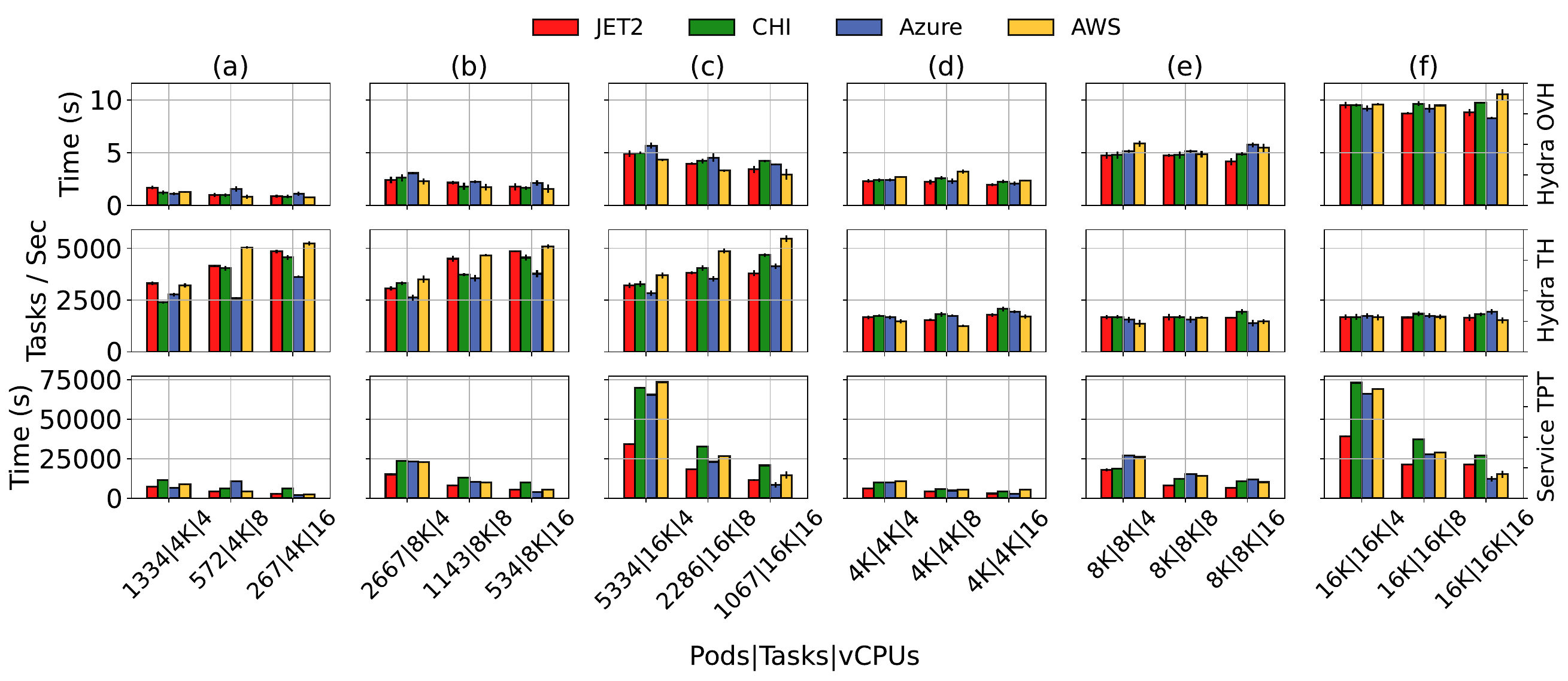}
    \caption{Weak and strong scaling of Hydra's OVH (top) and TH (middle) and cloud provider TPT (bottom). Measured on Jetstream2, Chameleon, Azure, and AWS with MCPP (a, b, c) and SCPP (d, e, f). Weak scaling: 4K/4, 8K/8, 16K/16 tasks/vCPUs; strong scaling: 4K/[4,8,16] 8K/[4,8,16], 16K/[4,8,16] tasks/vCPUs.}\label{fig:exp1_all}
\end{figure*}

Our experiments measure four metrics: Hydra's overheads (OVH), Hydra's throughput (TH), task total processing time (TPT), and task total execution time (TTX). OVH measures the time spent by Hydra to prepare the workload for execution and to communicate with the platform middleware to initiate the workload execution. TH measures Hydra's throughput as tasks processed per second and not the number of tasks a provider executes per second, as the latter is independent of Hydra's design. TPT measures the time taken to execute the workload tasks and to prepare and shut down the task execution environments. As such, TPT measures the performance of a target platform and its runtime services when driven by Hydra. TTX measures the total time the target platform/service takes to execute all the tasks submitted by Hydra.

Table~\ref{table:exp_setup} details the four experiments we designed to study Hydra's performance and compare the individual and aggregated performance of clouds and HPC platforms. Experiments 1, 2, and 3 compare Hydra's strong and weak scaling when executing tasks on single cloud platforms (Experiment 1), multiple concurrent cloud platforms (Experiment 2), and multiple concurrent cloud and HPC platforms (Experiments 3 and 4).

As experiments 1, 2, and 3A focus on Hydra's performance, we use \texttt{noop} tasks, i.e., tasks with zero execution time. That allows us to isolate and measure only Hydra's overheads and the time taken by each platform to set up and shut down the task execution environment (i.e., TPT). TPT allows us to compare the magnitude of Hydra's overheads to those of the target platforms. Further, it also provides users---like those of FACTS---with a baseline performance on which to base brokerage choices.

With experiments 1, 2, and 3A we study Hydra's performance with two task partitioning models: Single-Container-Per-Pod (SCPP), where each container has its own resources, and Multiple-Containers-Per-Pod (MCPP), where each container shares the resources with other containers from within the same pod. SCPP supports applications where a subset of tasks can run independently. In contrast, MCPP supports applications in which tasks have runtime dependencies, allowing them to execute within the same pod concurrently.

Experiments 3B and 4 measure Hydra's performance with heterogeneous tasks, i.e., tasks requiring different amounts of resources and/or execution time, including the FACTS workflow. That helps to understand whether heterogeneity introduces appreciable differences in Hydra's overheads or throughput. In these experiments, we measure the workloads TTX instead of TPT as the latter would not change compared to what experiments 1--3 already measure. TTX offers comparative insight into the actual performance of different platforms and a baseline for future orchestration policies.

We used uniform VMs across cloud providers with the same number of vCPUs and a comparable amount of memory. Importantly, Chameleon and AWS offered Intel Haswell and Xeon virtual cores, respectively, while Jetstream2 offered AMD EPYC-Milan physical cores. On Bridges2, each node provided 128 AMD EPYC physical cores. As middleware, we used a Kubernetes cluster with between 1 and 16 nodes, deployed via Elastic Kubernetes Service (EKS) on AWS, Azure Kubernetes Services (AKS) on Azure, and a custom image on the NSF machines. On the HPC platform, we used RADICAL-Pilot.

\subsection{Experiment 1: Per Provider Scalability}\label{ssec:exp1}

For each cloud provider---Jetstream2, Chameleon, Azure, and AWS---we execute 4000, 8000, and 16,000 tasks on 4, 8, and 16 vCPUs. We partition those tasks between 267 and 16,000 pods, measuring strong and weak scaling for both MCCP and SCPP application scenarios.

Fig.~\ref{fig:exp1_all} (top) shows strong and weak scaling of Hydra's overheads (OVH) for MCPP (a, b, c) and SCPP (d, e, f). Weak scaling is slightly sublinear, while strong scaling is mostly linear across the providers. The number of tasks and pods dominates OVH, which is invariant across providers, and the number of vCPUs. That validates the separation of concerns in Hydra's design between broker and platform capabilities, i.e., the performance of the former is independent of the latter. SCPP OVH is, on average, $\sim$46\% larger than the MCPP one due to the increased number of I/O operations needed to partition, prepare, and serialize each pod.

Fig.~\ref{fig:exp1_all} (middle) shows two differences in Hydra's TH with MCPP and SCPP: (1) MCPP is, on average, $\sim$44\% higher than SCPP across providers and scale; and (2) while TH with SCPP is invariant across providers and the number of tasks and pods, TH with MCPP increases with the ratio of pods and tasks. The lower TH with SCPP compared to MCPP is due to the increase of OVH shown in Fig.~\ref{fig:exp1_all} (middle). Hydra takes more time with SCPP to process pods (i.e., more I/O operations), hindering its throughput. Similarly, TH with MCPP improves by reducing the number of pods.

Building and managing the pods in memory instead of writing them on disk can significantly reduce Hydra's I/O bottleneck, thus increasing its TH. This can be done by using Kubernetes Python API as an alternative to communicate with the Kubernetes cluster on the NSF and Commercial clusters. Further, a better implementation of the task partitioning algorithm can reduce the observed reduction of throughput observed with MCCP.

Fig.~\ref{fig:exp1_all} (bottom) shows that TPT consistently scales across all runs with small error bars. Jetstream2 performs better than Chameleon and AWS, but Azure scales better, consistently outperforming Jetstream2 with 16 vCPUs. Jetstream2 performance is due to the pinning of vCPUs to physical cores, compared to the pinning to threads on all the other platforms. Hypervisor optimizations explain Azure's better scaling, while Chameleon shows the worst scaling, likely due to a less optimized hypervisor than AWS and Azure. SCPP shows a $\sim$9\% increase compared to MCPP across all providers. That is due to the larger overheads of per-pod initialization, scheduling, and termination. Finally, note that Hydra OVH is marginal compared to the TPT, confirming that platform overheads are dominant over Hydra's.

\subsection{Experiment 2: Cross Provider Scalability}\label{ssec:exp2}

We measure Hydra's scaling behavior when concurrently preparing and managing the execution of 16,000, 32,000, and 64,000 tasks on four VMs, one for each cloud provider. We use the largest VMs with 16 vCPUs available on all four providers and divide the workload tasks across each VM equally (with our allocation the largest VM on Jetstream2 and Chameleon have 16 vCPUs). Furthermore, we measure the aggregated time for OVH, TH, and TPT with MCPP and SCPP and compare them to the results of Experiment 1. The goal is to assess the consistency between the two experiments to verify whether Hydra concurrency introduces additional overheads. We sample a relevant subset of runs of Experiment 1 to avoid unnecessary experimental duplications.

\begin{figure}

	\centering
	\includegraphics[width=0.75\textwidth]{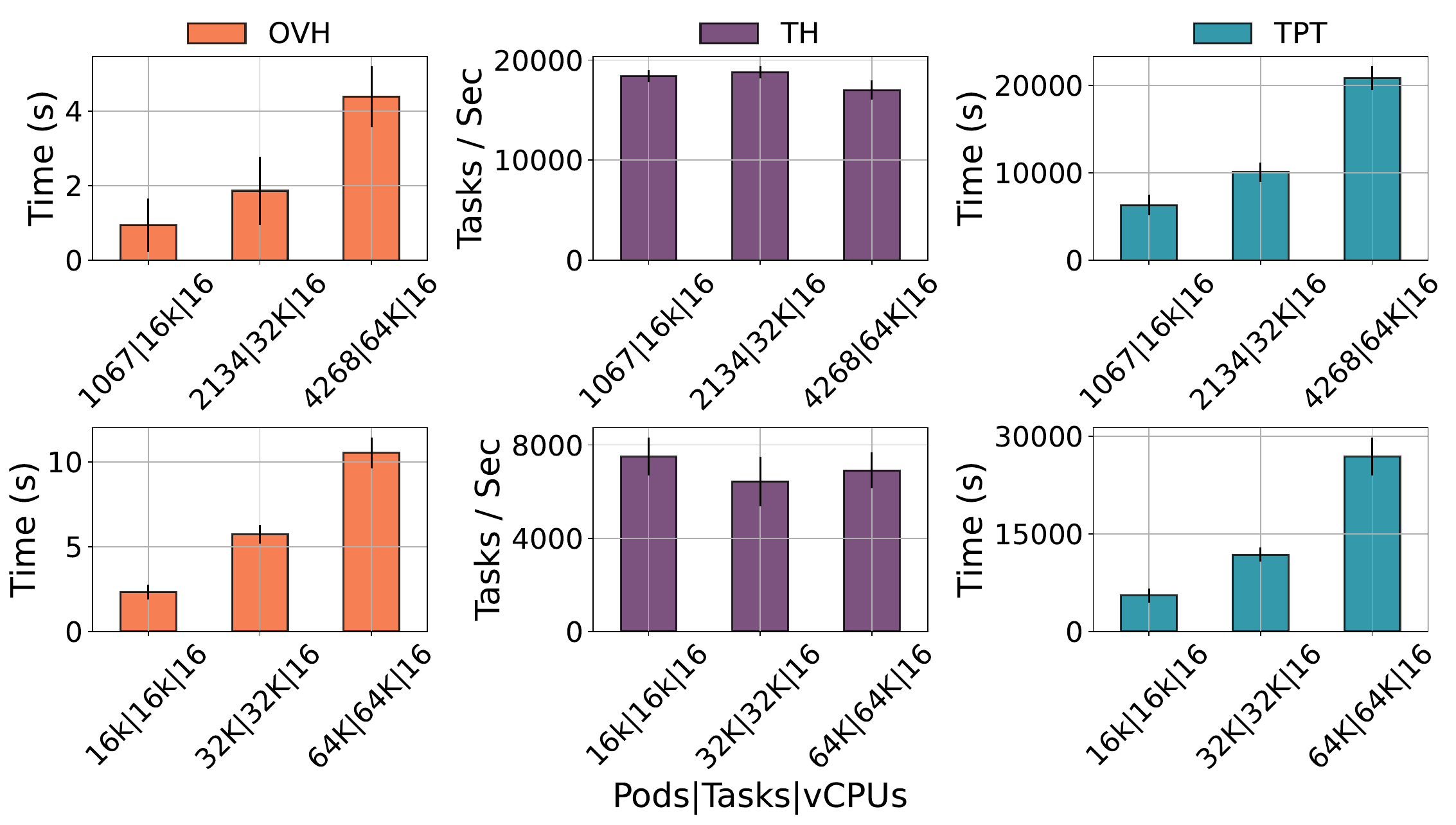}

	\caption{Aggregated TPT, OVH and TH on CHI, JET2, AWS, and Azure with MCPP (top) and SCPP (bottom).}\label{fig:nsf_com_ttx_ovh_ts_conc}

\end{figure}

Fig.~\ref{fig:nsf_com_ttx_ovh_ts_conc} shows that Hydra's aggregated OVH (orange) is consistent with the OVH measured on each provider in Experiment 1. Concurrently executing 16,000 tasks produces, on average, the same OVH measured running 4000 tasks on any of the four providers in Experiment 1. Experiment 2 data are consistent across the measured task and pod scales, replicating the already observed behavior with MCPP and SCPP. Hydra's aggregated TH (purple) is almost 4 times higher than the one measured in Experiment 1, confirming that Hydra can effectively and efficiently scale task throughput across diverse cloud providers. TH behavior is consistent with the one measured in Experiment 1, confirming that SCPP TH is lower than MCPP TH due to the increased cost of pod serialization and I/O. As expected, TPT (teal) is consistent with the provider performance measured in Experiment 1.

\subsection{Experiment 3: Cross Platform Scalability}\label{ssec:exp3}

We introduce three types of heterogeneity: platform-, node- and task-level. Hydra's core capabilities are to enable brokerage and execution management/monitoring across private/commercial cloud providers and HPC platforms. Experiment 3A measures Hydra's scalability with platform heterogeneity, while Experiment 3B also introduces node and task heterogeneity. We use only SCPP as it best fits a scenario where tasks execute outside a pod on HPC resources.

\begin{figure}
	\centering
	\includegraphics[width=0.75\textwidth]{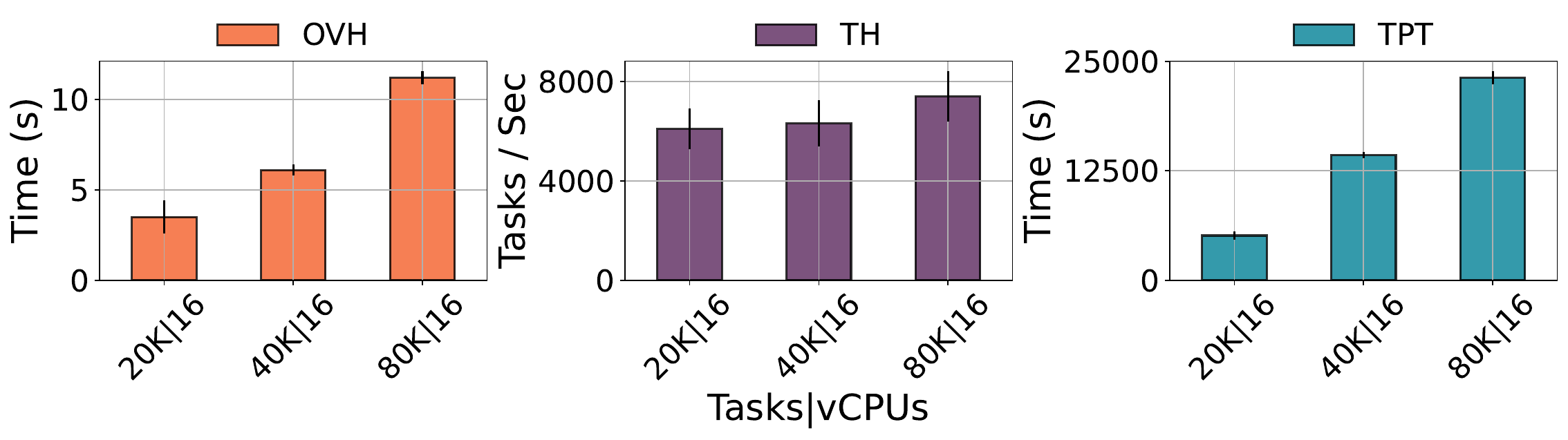}
	\includegraphics[width=0.75\textwidth]{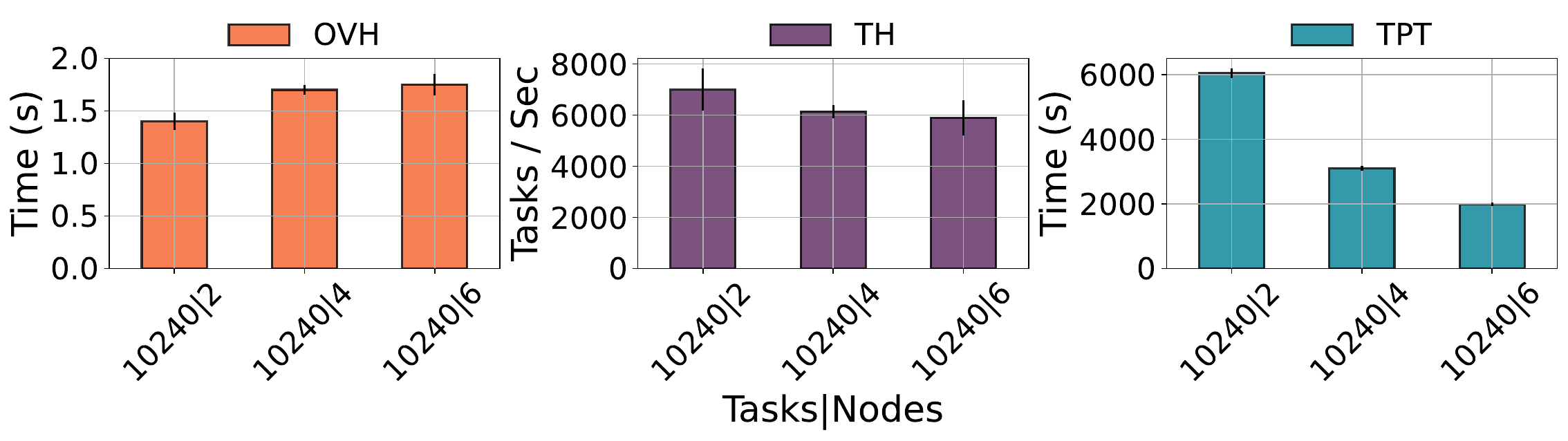}

	\caption{Aggregated TPT, OVH, and TH on four cloud providers (CHI, JET2, AWS,
	Azure) and ACCESS Bridges2 HPC platform with homogeneous (top) and heterogeneous (bottom)
	workloads and resources.}
	\label{fig:hpc_nsf_com_ttx_ovh_ts_conc}

\end{figure}

Experiment 3A setup is consistent with Experiments 1 and 2 to enable comparison (see Table~\ref{table:exp_setup}). Considering the error bars and the slight increase in the number of tasks of Experiment 3A, Fig.~\ref{fig:hpc_nsf_com_ttx_ovh_ts_conc} (top) shows that Hydra's OVH and TH are similar to those measured in Experiment 2 with SCPP. Thus, HPC-specific capabilities implemented in Hydra do not introduce overheads that are more significant than those required for cloud providers. TPT is also comparable to those of Experiment 2. However, it is important to note that we experienced short and consistent queuing time across all the experiment runs on the HPC platform. With a higher and less uniform queuing time, the aggregated TPT of Experiment 3A would increase compared to using only cloud platforms.

Experiment 3B represents scientific workloads in which different task types must execute with different degrees of concurrency. Supporting those workloads is particularly relevant for AI/ML-enabled workflows in which learning, inference, and simulation tasks often coexist within a single application~\cite{ward2023cloud,pascuzzi2023asynchronous}. Experiment 3B executes tasks with different durations and sizes on multi-node Kubernetes clusters and multiple HPC compute nodes. Tasks execute for 1--10 seconds on 1--4 CPUs and 0--8 GPUs. Short durations and relatively small sizes stress Hydra's performance, offering a `worse case' scenario for Hydra's performance characterization.

Fig.~\ref{fig:hpc_nsf_com_ttx_ovh_ts_conc} (bottom) shows that Hydra's OVH increases $\sim$5\% above two nodes but remains very similar between 4 and 6, confirming that adding nodes introduces only marginal overheads and that the number of tasks and pods remain dominant. Accounting for error bars, TH remains essentially invariant across the number of nodes. At the same time, TPT scales linearly between 2 and 4 nodes and sublinearly between 4 and 6 nodes, primarily due to increased Kubernetes overheads.

\subsection{Experiment 4: Use Case Scalability}\label{ssec:exp4}

We study the scalability of Hydra when managing the concurrent execution of the FACTS workflow (See \S\ref{sec:usecases}) on Jetstream2, AWS, and Bridges2. We characterize strong and weak scaling, measuring FACTS TTX (blue, orange, green) and Hydra OVH (red). Unlike the other experiments, Hydra has to deploy a stack on both cloud and HPC platforms that enables the execution of workflows, not just workloads (i.e., a set of independent tasks). Hydra deploys a multi-node Kubernetes cluster on the cloud platforms with the Argo workflow manager. In contrast, Hydra uses RADICAL-EnTK and RADICAL-Pilot on the HPC platform to execute the FACTS workflow~\cite{kopp2023framework}. Note that Hydra still has to manage the life-cycle of the deployed resources and the workflows on HPC and cloud concurrently.

We use Hydra's Python API to implement the FACTS workflow, which contains four main steps: pre-processing, fitting, projecting, and post-processing. Each step requires 1 core, 2GB of RAM, and the pre-processing and the fitting steps require input data files pre-staged on each target platform. FACTS workflow can be run multiple times to explore a vast problem space, and Hydra enables those workflow instances to be run on diverse platforms concurrently. Experiment 4 serves to validate the end-to-end capabilities of Hydra and to provide FACTS users with insight into which platform(s) offer better performance.

We run between 50 and 800 FACTS workflows on 16--256 AWS and Bridges2 cores and between 50 and 400 workflows on 16--128 cores on Jetstream2, as Jetstream2 has fewer cores compared to AWS and Bridges2. We maintain similar setup and resource properties of cores per node, RAM, and storage on cloud platforms. On HPC, we use 128 cores per node (compared to 16 cores on the clouds) as Bridges2 does not offer smaller compute nodes (see Table~\ref{table:exp_setup}).

\begin{figure}
	\centering
	\includegraphics[width=0.75\textwidth]{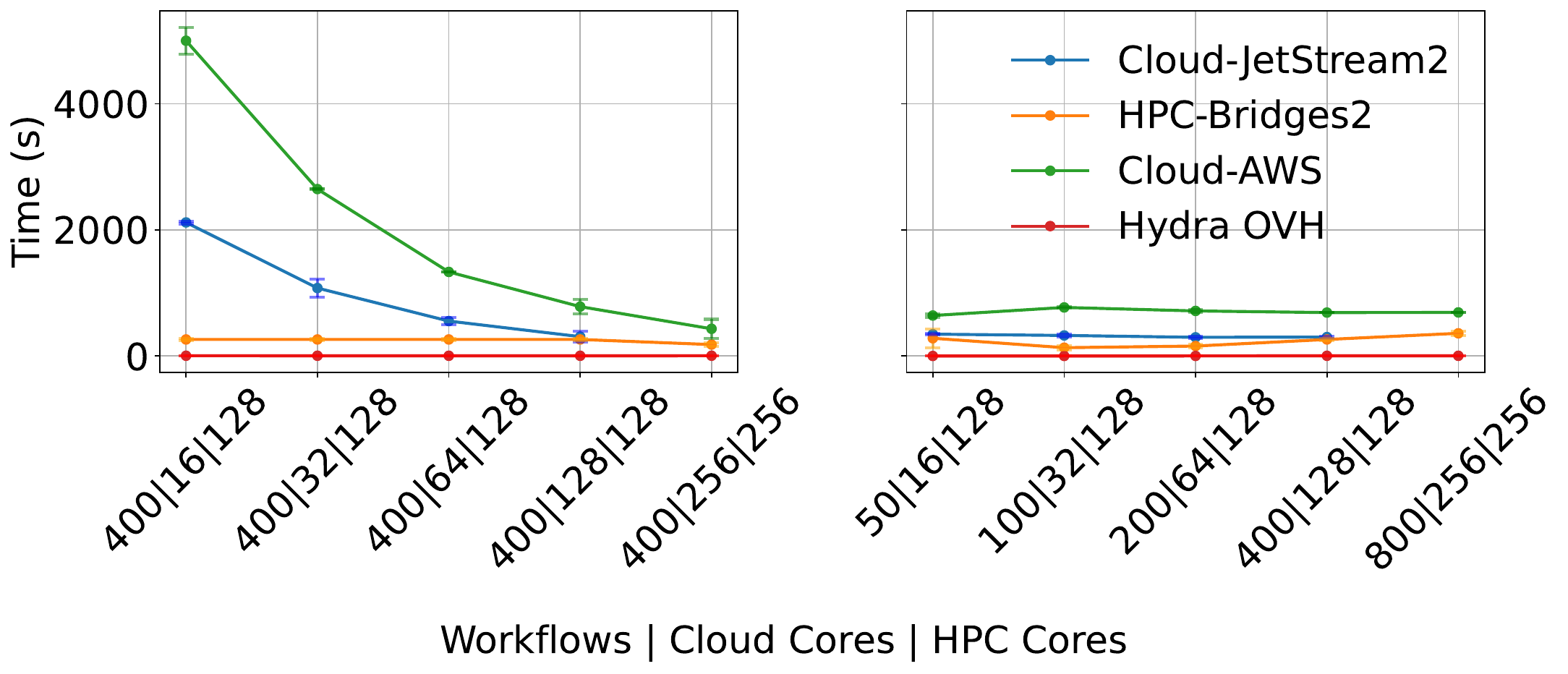}

	\caption{FACTS strong (right) and weak (left) scaling on Jetstream2 (blue), AWS (green) and Bridges2 (orange).}
	\label{fig:facts_exp}

\end{figure}

Fig.~\ref{fig:facts_exp} shows that for both strong (left) and weak (right) Hydra's OVH (red) scaling is consistent with what was observed in the previous experiments and, thus, invariant across workload and resource types. Further, OVH is negligible compared to the FACTS workflow makespan. With strong scaling, TTX on AWS (green) and Jetstream2 (blue) scale sublinearly due to the increasing platform overheads. Bridges2 (orange) has a linear behavior on 128 cores and scales with 256 cores. That is expected because Bridges2 does not allow acquiring less than 128 cores; thus, the first 4 runs of the experiment have the same concurrency. With weak scaling, TTX is close to the ideal scaling behavior on all the platforms.

Jetstream2 performs $\sim$2.5 times better than AWS due to the already mentioned different vCPUs mapping. Bridges2 performs $\sim$5 times better than Jetstream2 and $\sim$10 times better than AWS. That is because Bridges2 (1) has a higher number of cores per node, which allows for a greater level of concurrency, (2) does not suffer from virtualization overheads, and (3) has better CPU capabilities compared to AWS and Jetstream.

\section{Conclusions and Future Work}\label{sec:conclusions}

We introduced Hydra, a brokering system that manages, monitors, and provisions inter- and cross-cloud/HPC resources and manages, monitors, and traces the execution of heterogeneous workloads on those resources. Hydra's design is modular, and its interfaces allow integration with existing workflow middleware and runtime systems. That facilitates supporting new cloud providers and HPC platforms with a connector-based design.

Hydra addresses some critical challenges posed by resource management and workflow execution with resource, platform, task, and middleware heterogeneity. Specifically, Hydra enables concurrent use of NSF and commercial clouds alongside NSF and DOE HPC platforms. That satisfies the growing need of scientific communities to efficiently and effectively utilize a variety of clouds and HPC platforms in government-funded and commercial spaces. Finally, Hydra is designed from the ground up for performance, including tracing and monitoring capabilities.

Our experiments characterized Hydra's performance, showing minimal overheads compared to the overheads of private and commercial clouds and HPC platforms. Given its design (\S\ref{ssec:design}), Hydra's overheads are dominated by the number of tasks and pods processed but remain essentially invariant across cloud providers (Experiment 1, \S\ref{ssec:exp1}), HPC platforms (Experiment 2, \S\ref{ssec:exp2}) and type of tasks (Experiment 3A, \S\ref{ssec:exp3}). Hydra's throughput scales strongly and weakly and is independent of platforms (Experiments 1 and 2) and task heterogeneity (Experiment 3B).

Currently, Hydra generates pods and partitions tasks over those pods by relying on the file system. As seen in Experiment 1, that is inefficient and reduces Hydra's throughput, especially with SCPP\@. Early prototyping confirms that generating the pods and partitioning the tasks in memory reduces Hydra's overheads and increases its task throughput.

Experiments 1--3 show Hydra's capability to concurrently acquire cloud and HPC resources across diverse platforms via either single/multi-node Kubernetes clusters or a pilot system. Experiment 4 shows how that capability supports executing a scientific workflow at scale, with unprecedented concurrency, and with minimal overheads. In Experiment 4, Hydra concurrently managed the execution of 800 instances of the FACTS workflow on 4--16 nodes of Kubernetes clusters and 1 pilot job on an HPC platform. Note that Hydra's scaling is limited mainly by the capabilities of the middleware it uses on cloud and HPC platforms. For example, we could use a larger number of more dense VMs on AWS/Azure or use RADICAL-Pilot on a much larger HPC platform like Frontier~\cite{titov2023novel} without modifying Hydra's current code or hindering either AWS/Azure or RADICAL-Pilot performance.

Experiment 4 also highlighted the flexibility of Hydra's design in integrating within the existing ecosystem. While experiments 1--3 only required Kubernetes capabilities, executing FACTS required a workflow manager. Integrating Argo needed minimal development effort, and Hydra used its standard API without additional overhead. As seen in \S\ref{sec:usecases}, FACTS will soon require heterogeneous resources to run AI/ML-enabled workflows, and Hydra can already drive Kubernetes stacks to execute CPU/GPU and MPI task
One of Hydra's research lines is exploring the scheduling trade-offs across different stacks. For example, further research is needed to understand how to couple Hydra-level task partitioning and Argo- and Kubernetes-level task scheduling for metrics like workflow makespan and resource utilization.

Finally, our experiments offered valuable insights into the performance of the target cloud and HPC platforms. We could compare the overheads of commercial and NSF cloud resources and, as with FACTS, understand the performance trade-offs between cloud and HPC resources. Currently, that information enables Hydra's users to make binding decisions about tasks and resources before starting the execution of the workflow. Nonetheless, as part of the ongoing Hydra development, we use this experimental insight to develop, evaluate, and compare orchestration capabilities that will enable dynamic and adaptive binding of tasks to resources at runtime.

\section*{Acknowledgment}
This work is supported by NSF-2103986 (Rhapsody) and NSF-1931512 (RADICAL-Cybertools).

\bibliographystyle{unsrtnat}
\bibliography{references}

\end{document}